# Superconductivity in $Cu_xTiSe_2$

E. Morosan[1*], H. W. Zandbergen[2], B. S. Dennis[3], J. W. G. Bos[1],

Y. Onose[4], T. Klimczuk[1], A. P. Ramirez[3], N. P. Ong[4], and R. J. Cava[1]

[1]Department of Chemistry, Princeton University, Princeton, NJ 08540, USA

[2]National Centre for HREM, Department of Nanoscience,

Delft Institute of Technology, Al Delft, The Netherlands

[3]Bell Laboratories, Lucent Technologies, Murray Hill, NJ 07974, USA

[4]Department of Physics, Princeton University, Princeton, NJ 08540, USA

*e-mail: emorosan@princeton.edu.

**Charge density waves (CDWs) are periodic modulations of the conduction electron density in solids. They are collective states that arise from intrinsic instabilities often present in low dimensional electronic systems. The layered dichalcogenides are the most well-studied examples, with $TiSe_2$ one of the first CDW-bearing materials known. The competition between CDW and superconducting collective electronic states at low temperatures has long been held and explored, and yet no chemical system has been previously reported where finely controlled chemical tuning allows this competition to be studied in detail. Here we report how, upon controlled intercalation of $TiSe_2$ with Cu to yield $Cu_xTiSe_2$, the CDW transition is continuously suppressed, and a new superconducting state emerges near x = 0.04, with a maximum $T_c$ of 4.15 K found at x = 0.08. $Cu_xTiSe_2$ thus provides the first opportunity to study the CDW to Superconductivity transition in detail through an easily-controllable chemical parameter, and will provide new insights into the behavior of correlated electron systems.**

Charge density wave (CDW) transitions are a frequent occurrence in transition metal chalcogenides due to their low structural dimensionality. Layered $MX_2$ compounds and chain-based $MX_3$ compounds, where M is a group 4 or 5 metal and X = S, Se, or Te, are the best known examples [1-7]. These transitions arise to allow electronic systems to minimize their energy by removing electronic states at the Fermi level. This is achieved by introducing a new structural periodicity at the Fermi wave vector, inducing a band gap. Superconductivity and the CDW state are two very different cooperative electronic phenomena, and yet both occur due to Fermi surface instabilities and electron-phonon coupling. A number of CDW-bearing materials are also superconducting [8-13], and the idea that superconductivity and CDW states are competing electronic states at low temperatures is one of the fundamental concepts of condensed matter physics.

The use of control parameters to tune superconducting and CDW transition temperatures has been extensively explored. Pressure (in $NbSe_2$[14], $TaS_3$, $(TaSe_4)_2I$ and $NbSe_3$ [10]), or doping (in $TaS_2$[15]), or both (in $Lu_5Ir_4Si_{10}$[13]), have been used to reduce the CDW transition temperature and increase the transition temperature of an existing superconducting state, indicative of the competition between them.

Surprisingly, no system has yet been reported in which the emergence of a superconducting state after a charge density wave state has been suppressed via doping has been studied in detail: a transition that implies a deep connection between the two states, *i.e.,* that the same electrons are participating in both transitions. $TiSe_2$ was one of the first CDW-bearing compounds known, and is also one of the most frequently studied as the nature of its CDW transition has been controversial for decades. The CDW transition, at approximately 200 K, is to a state with a commensurate (2a,2a,2c) wavevector without an intermediate incommensurate phase [3, 16, 17]. The commensurate CDW wavevector and electronic structure calculations indicate that, unlike the case in most materials, the CDW in $TiSe_2$ is not driven by Fermi surface nesting. The normal state is presently believed to be either a semimetal or a semiconductor with a small indirect gap [3, 16, 18 - 22]. Recent photoemission experiments [23] concluded that the CDW transition in $TiSe_2$ is a transition from a small indirect gap, semiconductor normal state, into a state with a larger indirect gap at a slightly different location in the Brillouin zone. Although all agree that the CDW transition in TiSe$_2$ is not driven by conventional Fermi surface nesting, its cause remains controversial, involving a soft phonon and possibly electron-hole coupling or an "indirect" Jahn-Teller effect [23].

Here we report how, upon controlled intercalation of TiSe$_2$ with Cu to yield Cu$_x$TiSe$_2$, the CDW transition is continuously suppressed, and a superconducting state emerges near x = 0.04 with a maximum T$_c$ of 4.15 K at x = 0.08. A CDW-Superconductivity phase diagram as a function of doping is developed for Cu$_x$TiSe$_2$, analogous to the antiferromagnetism-superconductivity phase diagram found for the high temperature superconductors. The results indicate that Cu$_x$TiSe$_2$ provides the first opportunity to study the CDW-Superconductivity transition in detail through an easily-controllable chemical parameter. Such studies have been critical in understanding the behavior of other correlated electron systems, but so far have been lacking in CDW-superconductivity systems.

TiSe$_2$ is a trigonal symmetry, layered compound [24]. The Ti atoms are in octahedral coordination with Se, in TiSe$_2$ layers that, in the pure compound, are Van der Waals bonded to each other. As Cu atoms are added, they occupy positions between the TiSe$_2$ layers (Fig. 1a, inset). This results in a systematic expansion of the unit cell with Cu content in Cu$_x$TiSe$_2$, as evidenced by the lattice parameters shown in Fig. 1a. The expansion of the cell parameters is maintained up to x = 0.11. For higher Cu contents, both *a* and *c* remain unchanged from their value at x = 0.11. It can therefore be concluded that the solubility limit for Cu in TiSe$_2$ is x = 0.11 ± 0.01.

Of particular interest is the evolution of the charge density wave with Cu doping. Electron and X-ray diffraction studies of pure TiSe$_2$ at low temperatures show the presence of reflections corresponding to the basic trigonal structure and also the 2a, 2c superstructure reflections associated with the CDW state [3, 19]. Figure 1b shows an electron diffraction pattern of Cu$_{0.03}$TiSe$_2$ taken at approximately 120 K with the crystal tilted away from the [001] zone such that several higher order Laue zones, with reflections *hkl* (*l* = -1, 0, 1 and 2) are visible. The superreflections are only observed for *l* = 2n + 1, such that they are not visible in the zero order Laue zone in the [001] orientation (where *l* = 0). The 2a, 2c superstructure reflections, as is indicated in the figure, are clearly seen, as they are in TiSe$_2$. Therefore, the charge density wave is still present at 120 K at this composition. Significantly, the characteristic CDW wavevector is unchanged by doping. Apart from the 2a, 2c superstructure reflections, which are very sharp and occur only in the higher order zones where $l = 2n + 1$, more streaked reflections can be seen. Raising the temperature by approximately 20 K results in the disappearance of the 2a, 2c superstructure reflections. The more streaked reflections are still present above the CDW transition temperature and remain visible in diffraction patterns taken at room temperature. Furthermore, they are present in the diffraction pattern for $Cu_{0.08}TiSe_2$, the optimal superconducting composition, at 120K (Fig. 1c) and also in pure $TiSe_2$ at room temperature. The positions of these diffuse peaks are only in part the same as those of the 2a, 2c superstructure, as is shown in the overlays. The diffuse superreflections are not confined to a small band at $l = 2n + 1$ but are present everywhere, indicating that they are also streaked along c* (perpendicular to the $TiSe_2$ planes): the streaking appears to be continuous in that direction. It would be of interest to characterize the diffuse scattering as a function of temperature and composition and determine its origin. It is likely associated with the soft phonon believed to accompany the CDW transition [25].

Figure 2a shows the temperature dependence of the magnetic susceptibilities for $Cu_xTiSe_2$ over the range of Cu solubility. The normal state susceptibility (*e.g.* at 300 K) increases with Cu content. This suggests that the Cu doping introduces carriers into the conduction band in $TiSe_2$, increasing the electronic density of states and therefore the Pauli paramagnetism. This is further confirmed by specific heat measurements, described below. A drop in the susceptibility of pure $TiSe_2$ is seen as the temperature is lowered below the CDW transition at 200 K, consistent with the decrease in electronic density of states that occurs on opening a gap at the Fermi level (the susceptibility becomes negative because the core diamagnetism is larger than the Pauli contribution). Upon doping with increasing amounts of Cu, the CDW state in $Cu_xTiSe_2$ exists until x = 0.06, as seen in the drops in the susceptibilities. The susceptibility drop decreases with increasing Cu content, implying that fewer states are gapped at the CDW transition. The CDW transition temperatures can be determined from the on-sets of the susceptibility drops, and decrease continuously with increasing Cu content. For x = 0.06, the CDW transition, marked by a very small change in susceptibility, is reduced below 60 K and is no longer visible for higher x. The fact that local moment magnetism is not generated by Cu doping indicates that the intercalated Cu has a formal oxidation state of +1, a $3d^{10}$ electron configuration.

A systematic change in the transport properties of $Cu_xTiSe_2$ occurs on increasing x. The resistivity of our pure $TiSe_2$ (Fig. 2b) is very similar to that previously reported [3]: a broad maximum occurs around 150 K, with the ratio $\rho(150K)/\rho(300K) = 4$, comparable with that of the stoichiometric crystals [3]. However, unlike the single crystals where the ratio $\rho(300K)/\rho(6K)$ is 3- 4, in our sample this ratio is smaller than unity, likely due to the fact that it is a polycrystalline pellet. As Fig. 2b shows, the resistivity maximum in $Cu_xTiSe_2$ associated with the CDW state broadens and moves towards lower temperatures with increasing Cu doping, until it becomes unobserved for x > 0.06. In addition, the overall resistivities decrease as x increases from 0 to 0.08, and the high temperature curves become metallic and linear in T. Beyond x = 0.08, the normal state resistivity is almost unchanged. If the ratio $\rho(300K)/\rho(6K)$ is used to follow the change in transport properties across the series, it can be seen that the smallest amount (x = 0.01) of Cu added to $TiSe_2$ yields a four-fold increase in the metallicity. The inset in Fig. 2b illustrates the metal-like behavior of the Seebeck coefficient (S) in a subset of the $Cu_xTiSe_2$ compounds. A sharp drop in S for pure $TiSe_2$ around 200 K marks the CDW transition, which is also associated with an apparent change of dominant carrier type (S changes sign). With Cu introduced to this material, the CDW transition is suppressed, and the Seebeck coefficient becomes negative for the whole temperature range shown. For all higher Cu contents, S is negative between 100 and 400 K, and, for example, at the optimal superconducting composition of x = 0.08, |S| decreases linearly with T between 400 K and 100 K, as expected for a metallic compound. The Seebeck data are thus consistent with electron doping of $TiSe_2$ to a metallic state as Cu is intercalated. Thus the susceptibility and transport data indicate that the CDW-metallic transition in $Cu_xTiSe_2$ is likely due to electron doping away from the ideal value required for CDW formation in $TiSe_2$.

As the $Cu_xTiSe_2$ compounds evolve into better metals, and the CDW state is suppressed with increasing Cu content, a superconducting state emerges at low temperatures for x > 0.04. This is illustrated in the low temperature magnetization and resistivity data shown in Fig. 3. Below x = 0.04, no superconducting transition is observed down to the lowest temperature of our measurements, T = 0.4 K. At x = 0.04 however, a drop in the resistivity just begins to appear as temperature is lowered to 0.4 K (Fig. 3b), suggesting the onset of a superconducting transition at that temperature. The superconducting transition is clearly observed for x = 0.045 and x = 0.05 and comes into the temperature range of the susceptibility measurements at x = 0.055 (Fig. 3a). $T_c$ reaches a maximum of about 4.15 K for x = 0.08 and then decreases for higher Cu contents (*e.g.* to 2.5 K for x = 0.10). Thus an optimal composition for superconductivity of $Cu_{0.08}TiSe_2$ is observed.

Specific heat measurements performed at H = 0, 0.25 T, 1 T and 2 T are shown in Fig. 3c as $C_p/T$ *vs.* $T^2$. As expected, the H = 0 specific heat data (full circles) display a peak at the superconducting transition temperature $T_c$ = 4.1 K, which moves down in temperature as magnetic field is applied. The normal state specific heat can be approximated, at low temperatures, as $C_p = \gamma T + BT^3$, where the former term represents the normal state electron contribution and the latter the lattice contribution, to the specific heat. When plotted as $C_p/T$ vs. $T^2$, the data in Fig. 3c is linear above the transition up to 6 K, and the extrapolation to T = 0 gives $\gamma$ = 4.3 mJ/mol $K^2$. The small $\gamma$ value and the corresponding superconducting transition temperature $T_c$ = 4.1 K place this compound in the $T_c$ *vs.* $\gamma$ regime of conventional superconductors [26]. BCS theory [27] predicts that, at the transition temperature, $\Delta C_e(T_c)/\gamma T_c$ = 1.49, and for $Cu_{0.08}TiSe_2$ this ratio, obtained from the entropy conservation construction in the inset in Fig. 3c, is approximately 1.68.

Detailed measurements of the field-dependence of the magnetization and the resistive transition of $Cu_{0.08}TiSe_2$, presented in Fig. 4, allow for a better characterization of the superconducting state. The magnetization isotherms for T = 1.8 K - 4.3 K (Fig. 4a) display typical type-II superconductor behavior. Together with the magnetoresistance data (Fig. 4b), these measurements yield the upper critical field values $H_{c2}$ shown as circles in Fig. 4c. The full circles in Fig. 4c represent the field values where the magnetization becomes 0, and the open circles correspond to the resistance on-sets on the $\rho(H)$ curves. Estimates of the lower critical field $H_{c1}$ have been determined from the magnetization data and thus are limited to temperatures above 1.8 K: as the inset in Fig. 4a shows, the magnetization is linear in field at low H values; $H_{c1}$ is estimated as the field values where departures from linearity occurred at each temperature. The anticipated linear temperature dependence close to $T_c$ is evident for both $H_{c1}$ and $H_{c2}$ (Fig. 4c), which also results in a linear thermodynamic critical field $H_c$ (Fig. 4c), calculated as $H_c = \sqrt{(H_{c1} \cdot H_{c2})}$.

Close to T = 0, BCS theory predicts that the upper critical field decreases with temperature as $H_c(T) \approx H_c(0)*(1 - 1.07*T^2/T_c^2)$ [27]. The dashed line in Fig. 4c represents a fit of the $H_{c2}$ data to this expression, yielding a $H_{c2}(0)$ value of approximately 1.39 T. (The zero-field $T_c$ (3.2K) estimated from this fit is smaller than the measured value of 4.15 K, but the value of $H_{c2}(0)$ is well defined.) The high temperature data (close to $T_c$) can also be used to estimate $H_{c2}$, based on the equation $H_{c2}(0) = 0.693 \cdot H^*_{c2}(0)$ [28], where $H^*_{c2}(0) = - (dH_{c2}/dT)_{Tc} \cdot T_c$. The dotted line in Fig. 4c represents the extrapolation to T = 0 of the linear fit at high temperatures, yielding an estimate for $H_{c2}(0) = 1.27$ T. Based on these measurements, we conclude that the upper critical field $H_{c2}(0)$ of $Cu_{0.08}TiSe_2$ is $H_{c2}(0) = (1.33 \pm 0.06)$ T. Despite displaying the expected quadratic temperature dependence at low temperatures, the $H_{c2}$ values determined from M(*H*) measurements are likely overestimates of the actual values. This could be a result of using polycrystalline pellets rather than single crystals, particularly if the critical field is anisotropic: the polycrystalline samples yield an average value $H_{c2}$ that is intermediate between the values corresponding to H||ab and H||c. Using the above critical field values to estimate the Ginzburg-Landau parameter $\kappa = \lambda/\xi \approx H_{c2}/H_{c1}$, it can be concluded that $Cu_{0.08}TiSe_2$ is in the extreme type-II limit, as $\kappa \approx 1T/0.01T = 100$.

The variation of the transport, magnetic, and thermodynamic qualities of the normal state in the $Cu_xTiSe_2$ series is summarized in Fig. 5. The resistivity, specific heat, magnetic susceptibility, and Seebeck coefficient data taken together indicate that the Cu atoms contribute electrons to the conduction band on doping. This electron doping suppresses the CDW and induces metallic behavior in $Cu_xTiSe_2$ with a resistivity near $10^{-4}$ Ω-cm at room temperature in the metallic phase. As the carriers are introduced, the electronic contribution to the specific heat, γ, increases from approximately 1 to approximately 4 mJ/mol K$^2$ at the optimal superconducting composition. Estimates of the Wilson ratio R = $\chi_0/(3\gamma)(\pi k_B/\mu_B)^2$ based on the measured susceptibilities alone yield values between 0.3 and 0.4, much smaller than the expected R = 1 value for the free electron approximation. The small susceptibilities observed in this system must be corrected for core diamagnetism; however, correcting for the core contributions of Ti 4+ and Se 2- [29], results in R values that appear to be too high (between 2.5 and 5). This suggests that additional contributions to the observed susceptibility need to be considered to fully understand this system. This analysis is left to a future study.

Finally, the overall behavior of this system is summarized in the electronic phase diagram presented in Fig. 6. Using Cu doping as a finely controlled tuning parameter, the CDW transition in $TiSe_2$ is driven down in temperature, and a new superconducting state emerges. The superconducting state appears for $x > 0.04$, going through a maximum $T_c$ of 4.15 K at $x = 0.08$, followed by a decrease of $T_c$ before the chemical phase boundary is reached at $x = 0.11$. There is a small boundary composition region ($0.04 < x < 0.06$) where superconductivity and CDW behavior appear to coexist. The reason why superconductivity arises from the CDW state in $TiSe_2$ upon Cu doping has not yet been determined. It may be that Cu doping results in a tendency towards increasing the dimensionality of the Fermi surface, destabilizing the CDW and allowing for correlations to build in a third dimension, tipping the balance in favor of superconductivity. Otherwise, the superconductivity may emerge from the CDW state due to the change in electron count upon Cu doping. Further study of $Cu_xTiSe_2$ will be of interest to determine which of these is the underlying cause of the transition between the competing CDW and superconducting states, as will additional detailed and generic studies made possible by the fine chemical tuning of the electronic system that $Cu_xTiSe_2$ affords.

Correspondence and requests for materials should be sent to E. Morosan (emorosan@princeton.edu) or R. J. Cava (rcava@princeton.edu).

**Acknowledgements**

This research was supported primarily by the US DOE-BES solid state chemistry program, and, in part, by the US NSF MRSEC program.

**Methods**

Polycrystalline $Cu_xTiSe_2$ samples ($0 \leq x \leq 0.14$) were prepared in two steps. Firstly, stoichiometric amounts of elemental powders were sealed in evacuated silica tubes and heated from room temperature to 350ºC in about 1 hour. The temperature was then increased at 50ºC/hr to 650ºC, after which it was maintained at 650ºC for 20 hours. Secondly, the powders were pressed into pellets, re-sealed in silica tubes under vacuum, and annealed at 650ºC for 50 hours. Homogeneous, purple-grey pellets were obtained.

Powder X-ray diffraction measurements were employed to characterize the samples. Room temperature data were recorded on a Bruker D8 diffractometer, using Cu Kα radiation and a diffracted beam monochromator. Electron diffraction was performed with Philips CM300UT and CM30T electron microscopes, operated at 300 kV, using image plates as recording media. Electron transparent areas of the specimens were obtained by crushing slightly under ethanol to form a suspension, and then placing a droplet of this suspension on a carbon-coated holey film on a Cu or Au grid. For the cooling experiments, a Gatan liquid nitrogen cooling holder was used for the CM30T, and a home made one for the CM300UT. The Gatan holder allows for an estimation of the sample temperature, but it is measured far from the sample area, resulting in an underestimate of the sample temperature. Low temperature diffraction data were taken primarily at an indicated temperature of 87 K, but, because the temperature is measured away from the sample cap (as described above), we estimate the lowest temperature reached to be approximately 100 - 120 K. The heating of the sample due to the electron beam could also be a factor, but we excluded this as the major effect by measurements with low beam intensities.

Magnetization measurements as a function of temperature and applied field were performed in a Quantum Design MPMS SQUID magnetometer. Temperature-dependent resistivity measurements in constant applied fields were taken in a Quantum Design PPMS-9 instrument, using a standard four probe technique. Additional resistivity measurements down to T = 0.4 K were performed in a $^3$He refrigerator inserted in a 7 T superconducting magnet. Thermopower data were collected using an MMR technologies

SB100 Seebeck measurement system. Specific heat data were also collected in the Quantum Design PPMS-9 instrument, by the thermal relaxation method.

**Figure Captions**

**Fig.1** Lattice parameters of $Cu_xTiSe_2$. (a) Change in the lattice parameters of $Cu_xTiSe_2$ with Cu content x. The solid lines reflect the expected Vegards law scaling of a and c with x. Inset: the crystal structure of $Cu_xTiSe_2$. (b - c) Electron diffraction patterns of the $Cu_{0.03}TiSe_2$ and $Cu_{0.08}TiSe_2$ reciprocal lattices, with the 2a, 2c supercell outlined. The crystals are tilted away from the (001) zone axis, to show the superreflections in higher order Laue zones. The diffraction pattern in (b) shows sharp 2a, 2c superstructure reflections, with the unit cell indicated. The diffraction pattern in (c) shows diffuse superreflections (the unit cell is also shown); these streaked superreflections are also visible in (b), and are particularly obvious in areas where the 2a, 2c superstructure is absent.

**Fig.2** Magnetization and transport properties of $Cu_xTiSe_2$. (a) $Cu_xTiSe_2$ M(T) curves measured in a constant H = 0.5 T applied field, for 0 < x < 0.10. Solid lines illustrate how the CDW transition temperatures have been determined. (A small peak is seen around 60 K in a few of the measurements, which is attributed to an oxygen impurity trapped in the measurement system.) (b) H = 0 temperature-dependent resistivity data for 0 < x < 0.10; inset: Seebeck coefficient for x = 0, 0.01, 0.03 and 0.08.

**Fig.3** The superconducting phase transition as function of Cu content x. (a) H = 5 Oe AC magnetization data at low temperatures, showing the sharp superconducting transitions for x = 0.055, 0.06, 0.08 and 0.10. (b) Low temperature ρ(T) data in zero-field for x = 0.04, 0.045, 0.05 and 0.08. (c) Field-dependent specific heat data around the superconducting transition in $Cu_{0.08}TiSe_2$. The dotted line represents the extrapolation of the linear fit of $C_p/T$ ($T^2$) from ~ 6 K down to 0, yielding an electronic specific heat coefficient γ of 4.3 mJ $mol_{Ti}^{-1}$ $K^{-2}$. Inset: Electronic contribution to the specific heat of the superconducting (full symbols) and normal (open symbols) state of $Cu_{0.08}TiSe_2$. The solid line represents the entropy-conservation construction which gives the ratio $\Delta C_{es}/\gamma T_c$ to be ~1.68.

**Fig.4** Characterization of the superconductivity in $Cu_{0.08}TiSe_2$: (a) M(H) curves for T = 1.8 K, 2.0 K, 2.5 K, 3.0 K, 3.25 K, 3.5 K, 3.75 K, 4.0 K and 4.3 K, where red arrows mark the position of $H_{c2}(T)$; inset: the low-H part, with the position of $H_{c1}(T)$ marked by black arrows. (b) $\rho(H)$ data for T = 0.36 K, 0.6 K - 2.0 K ($\Delta T$ = 0.02 K), and 2.5 K - 4.5 K ($\Delta T$ = 0.05 K), with red arrows marking the $H_{c2}(T)$ values. (c) $H_c$ - T phase diagram, including $H_{c1}(T)$ (triangles) and $H_{c2}(T)$ (circles) determined from M(H) and $\rho(H)$ data, and the calculated thermodynamical critical field $H_c(T)$ calculated as $H_c=\sqrt{(H_{c1} \cdot H_{c2})}$ (crosses). As predicted by BCS theory, the solid lines represent linear fits around $T_c$ and the low-temperature dashed line is a fit to $H_{c2}(T) \approx H_{c2}(0) (1 - 1.07*T^2/T^2_c)$.

**Fig.5** Summary of the composition dependent properties in $Cu_xTiSe_2$: M/H(300K), $\rho(300K)/\rho(6K)$, electronic specific heat coefficient $\gamma$ and Seebeck coefficient S(300K) as a function of Cu composition x. The solid lines are guides to the eye showing the linear variation of the latter two quantities, whereas for the former two the lines reflect steep changes at low Cu content (0 < x < 0.02) and through the superconducting state (0.04 < x < 0.08).

**Fig.6** The $Cu_xTiSe_2$ T - x electronic phase diagram: open circles represent the CDW transition temperature, and the full-circles correspond to the superconducting transition temperature. The shaded circle at x = 0.04 indicates that the transition temperature is just below our minimum available temperature, and the dashed circle at x = 0.06 marks the barely visible CDW transition at x = 0.06.

Fig.1.

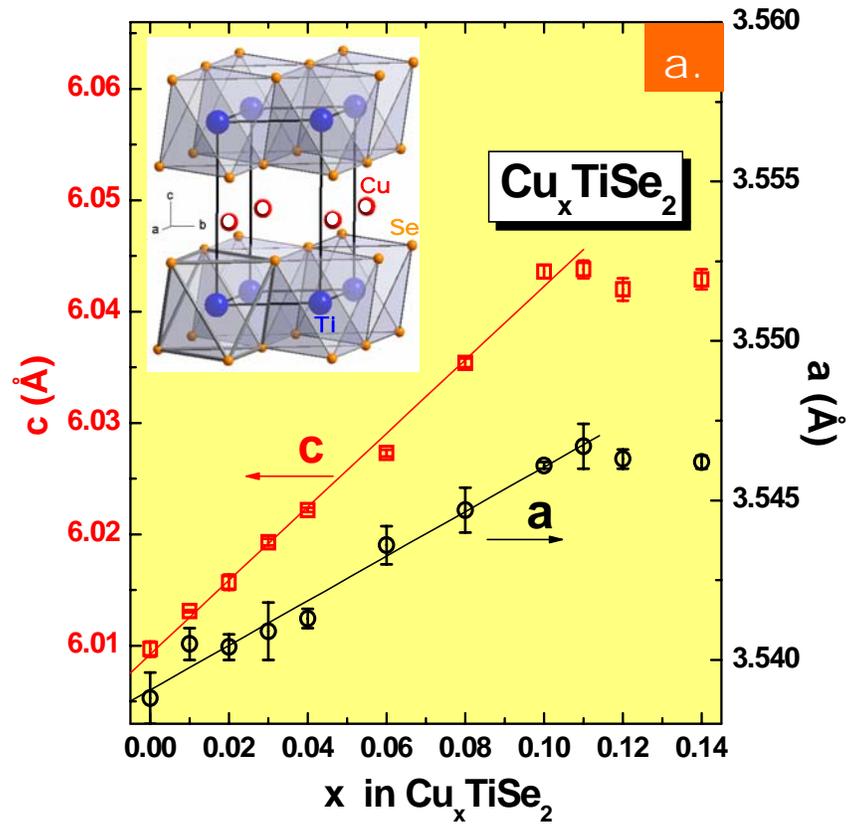
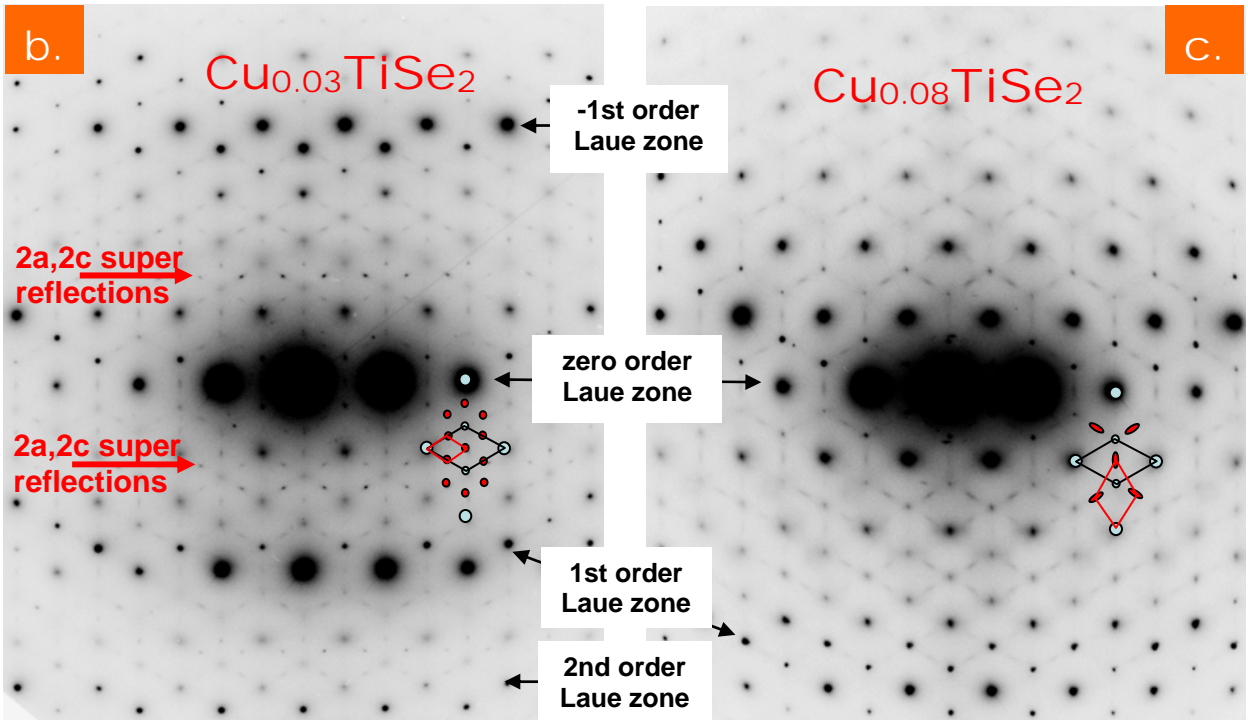

Fig.2.

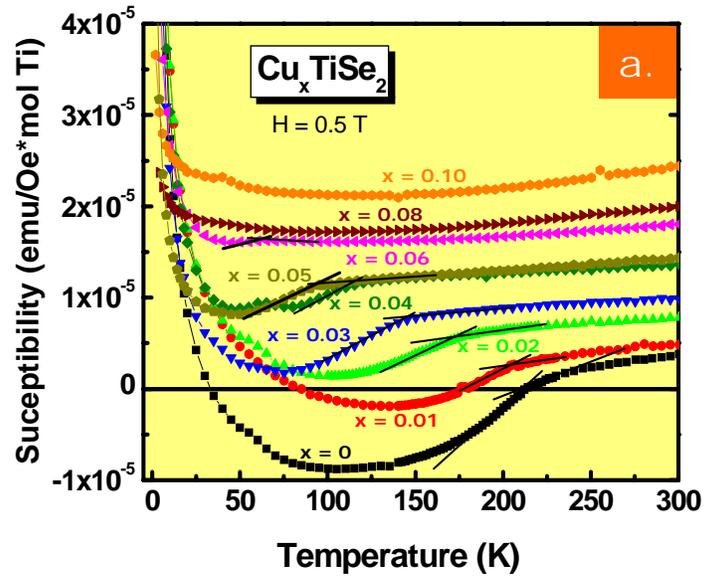

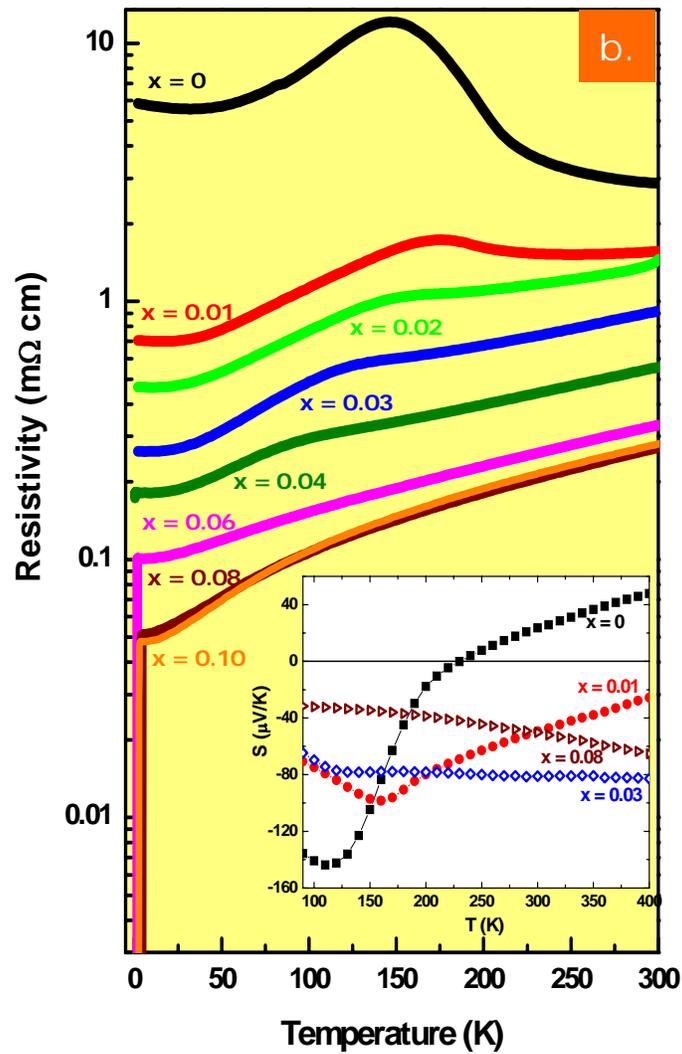

Fig.3.

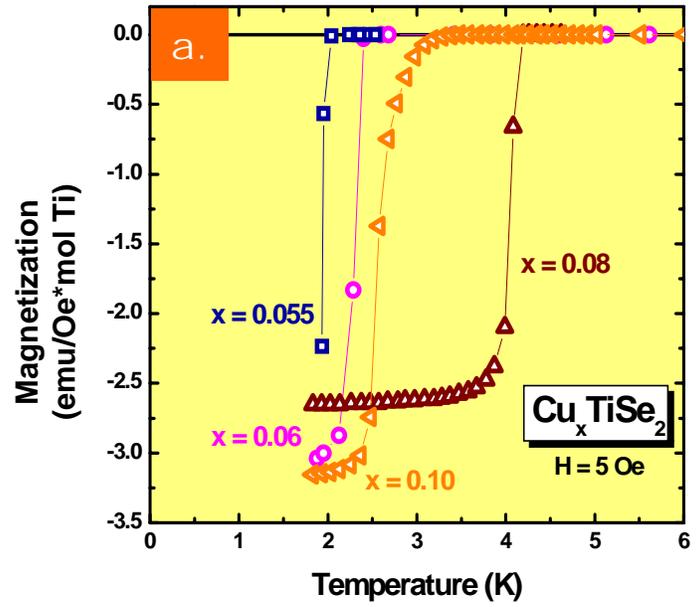

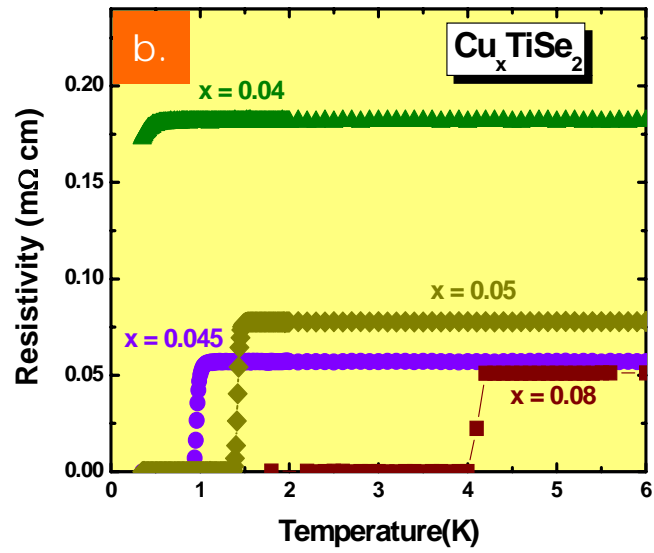

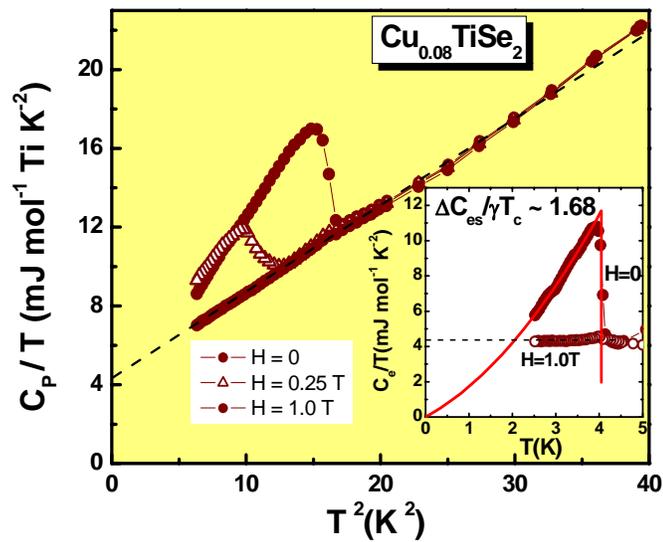

Fig.4.

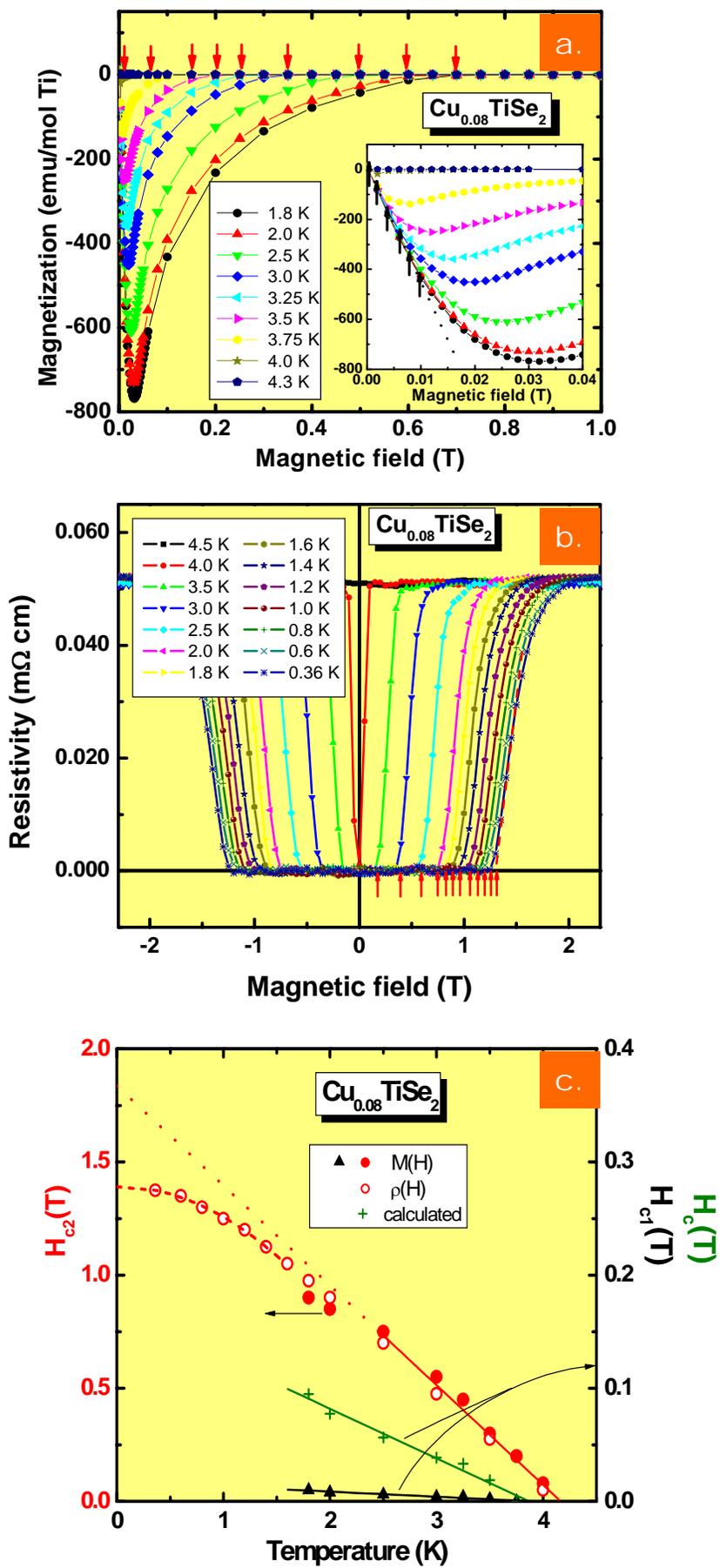

Fig.5.

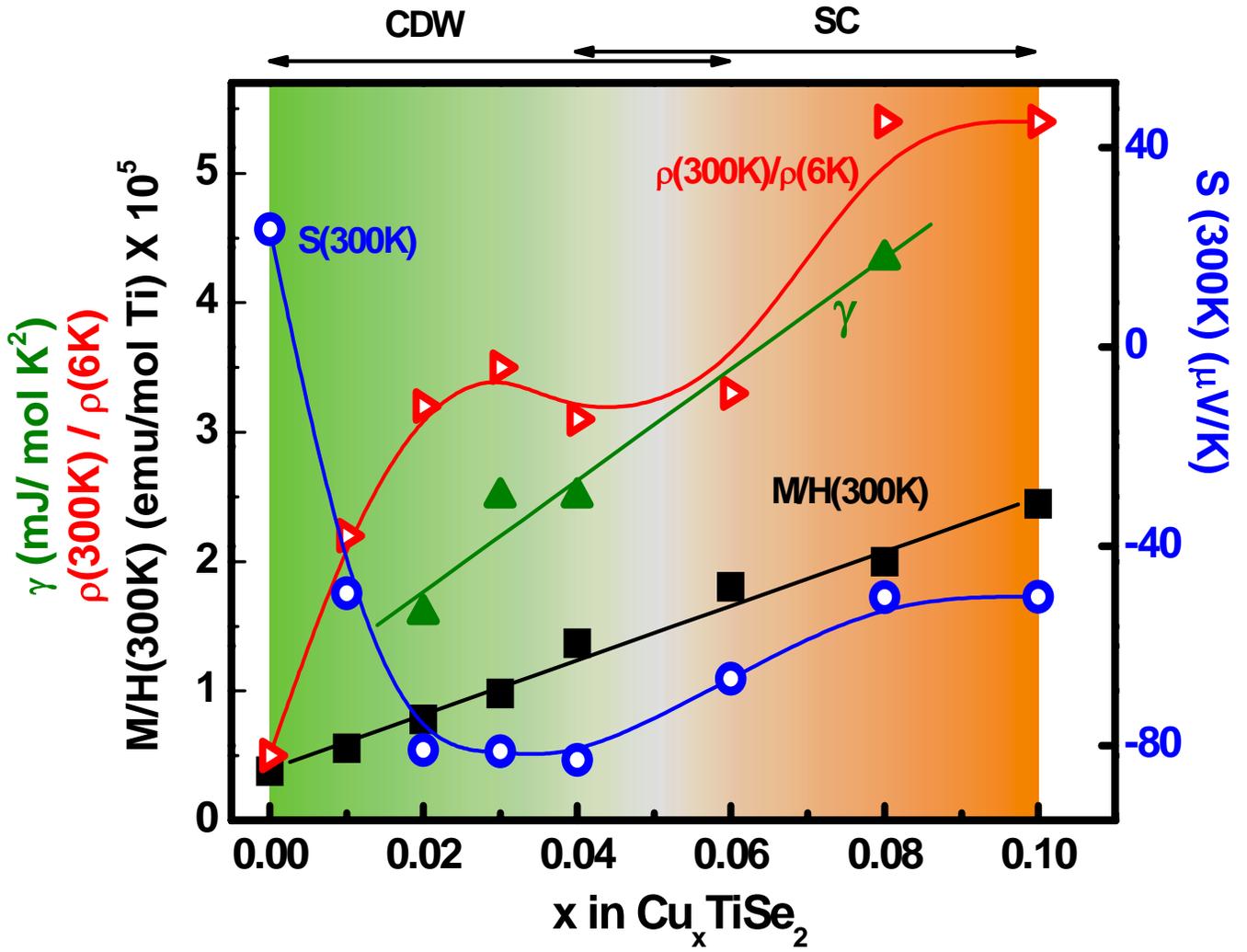

Fig.6.

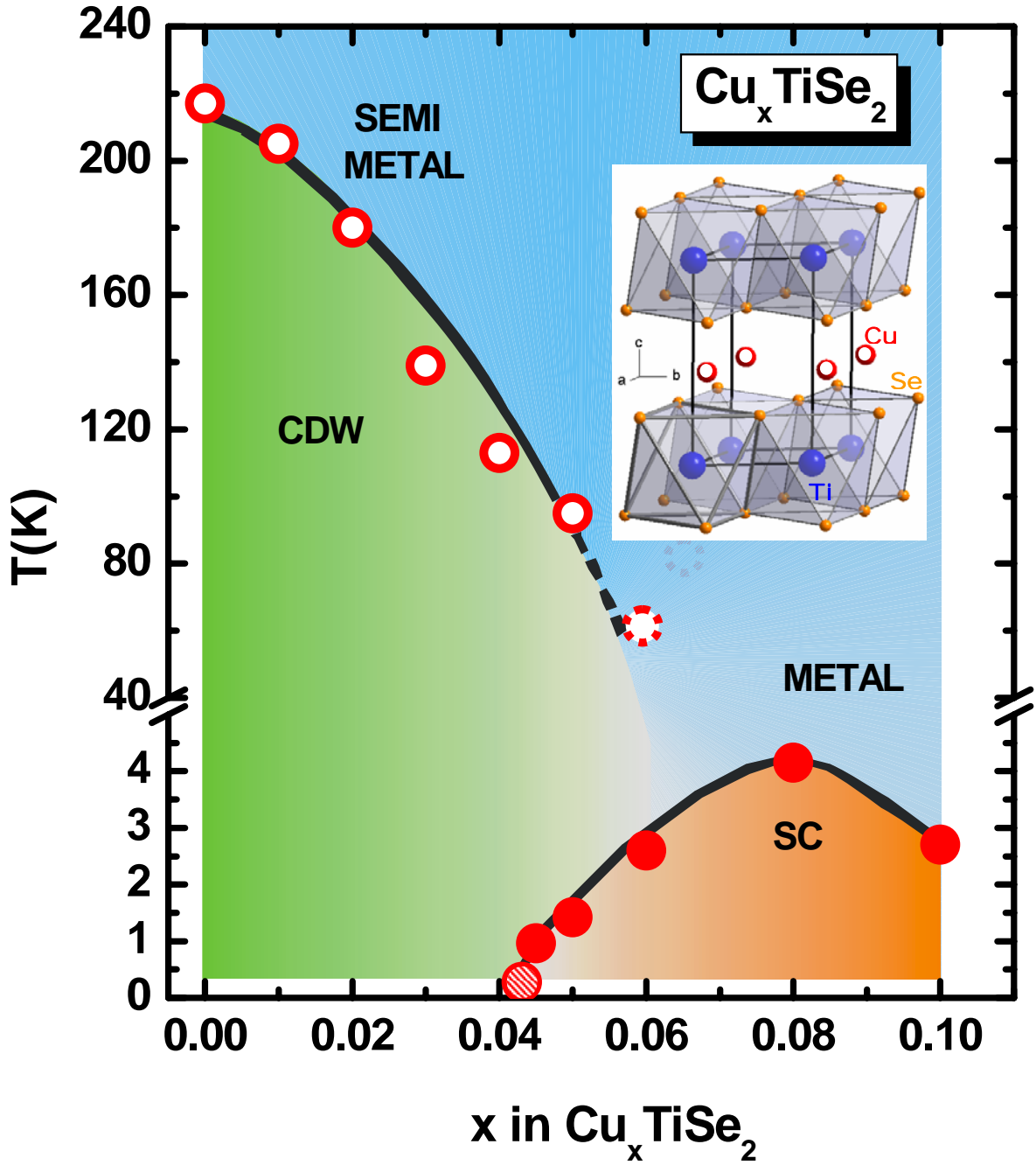